\documentclass[12pt]{article}

\usepackage{epsfig}

\usepackage{geometry}
\usepackage{caption}

\def\bphi{\mbox{\boldmath $\phi$}}

\def\bPhi{\mbox{\boldmath $\Phi$}}

\def\Tr{\rm Tr}

\def\tfrac#1#2{{\textstyle{{#1}\over{#2}}}}

\def\half{\tfrac{1}{2}}
\def\third{\tfrac{1}{3}}
\def\quart{\tfrac{1}{4}}

\def\twothirds{\tfrac{2}{3}}

\def\gsm{(g_1^{\rm SM})}

\begin{document}

\begin{titlepage}

\baselineskip 24pt

\begin{center}

{\Large {\bf The $Z$ boson in the Framed Standard Model}}

\vspace{.5cm}

\baselineskip 14pt

{\large Jos\'e BORDES \footnote{Work supported in part by  Spanish
    MINECO under grant FPA2017-84543-P,
Severo Ochoa Excellence Program under grant SEV-2014-0398, and
Generalitat Valenciana under grant GVPROMETEOII 2014-049.}}\\
jose.m.bordes\,@\,uv.es \\
{\it Departament Fisica Teorica and IFIC, Centro Mixto CSIC, Universitat de 
Valencia, Calle Dr. Moliner 50, E-46100 Burjassot (Valencia), 
Spain}\\
\vspace{.2cm}
{\large CHAN Hong-Mo}\\
hong-mo.chan\,@\,stfc.ac.uk \\
{\it Rutherford Appleton Laboratory,\\
  Chilton, Didcot, Oxon, OX11 0QX, United Kingdom}\\
\vspace{.2cm}
{\large TSOU Sheung Tsun}\\
tsou\,@\,maths.ox.ac.uk\\
{\it Mathematical Institute, University of Oxford,\\
Radcliffe Observatory Quarter, Woodstock Road, \\
Oxford, OX2 6GG, United Kingdom}

\end{center}

\vspace{.3cm}

\begin{abstract}

The framed standard model (FSM), constructed initially for explaining the
existence of three fermion generations and the hierarchical mass and mixing
patterns of quarks and leptons \cite{dfsm,tfsm}, suggests also a ``hidden sector''
of particles \cite{cfsm} including some dark matter candidates.  It predicts
in addition a new vector boson $G$, with mass of order TeV, which mixes with
the $\gamma$ and $Z$ of the standard model yielding deviations from the
standard mixing scheme, all calculable in terms of a single unknown
parameter
$m_G$.  Given that standard mixing has been tested already to great accuracy
by experiment, this could lead to contradictions, but it is shown here that for the
three crucial and testable
cases so far studied (i) $m_Z - m_W$, (ii) $\Gamma(Z \rightarrow
\ell^ + \ell^-)$, (iii) $\Gamma(Z \rightarrow$ hadrons), the deviations are
all
within the present stringent experimental bounds provided $m_G > 1$
TeV, but should soon be detectable if experimental accuracy
improves.
This comes about because of some subtle cancellations, which might have a
deeper reason that is not yet understood.  By virtue of mixing, $G$ can be
produced at the LHC and appear as a $\ell^+ \ell^-$ anomaly.  If found, it
will be of interest not only for its own sake but serve also as a window on
to the ``hidden sector" into which it will mostly decay, with dark matter
candidates as most likely products.

\end{abstract}

\end{titlepage}

\clearpage

\section{Introduction}

The framed standard model (FSM) is constructed from the standard model (SM) by 
adding to the usual gauge boson and matter fermion fields the frame vectors in 
internal gauge space as dynamical variables (called framons), thereby 
seemingly assigning 
a geometrical meaning to the Higgs field, suggesting an origin for the three 
fermion generations,  giving an explanation to the special patterns of fermion 
masses and mixing as seen in experiment, and providing a solution to the strong 
CP problem, as well as posing and answering some other questions of interest, 
all in terms of a small number of adjustable parameters \cite{dfsm}, 
\cite{tfsm}, \cite{cfsm}.

Though having thus apparently explained, as it was originally intended to do,
some outstanding peculiarities that the SM takes for granted as inputs from 
experiment, the FSM now faces the question whether the new framon degrees of 
freedom that it introduces might not lead to deviations from the SM beyond what 
is allowed by experiment.  Some of the resulting deviations are addressed in 
\cite{cfsm}, but the most acute and pressing is the following, which needs 
special attention.  The FSM predicts a new vector boson, say $G$, with mass of 
order TeV, which mixes with the photon and the $Z$ in the standard sector.  The 
urgent question is whether such  mixing would spoil the present, near-perfect, 
agreement with experiment of the standard model mixing scheme.

In brief (for details see  \cite{cfsm}), this situation comes about as
follows.  The framon as conceived in the FSM is a scalar field in two parts:
\begin{itemize}
\item the flavour framon:
\begin{equation}
\bphi  =  \left( \phi_r \right),\ r = 1, 2, 
\end{equation} 
and 
\item the colour framon: 
\begin{equation} \bPhi  =  \left( \phi_a^{\tilde{a}} \right),\
    a = 1, 2, 3; \  \tilde{a} = \tilde{1}, \tilde{2}, \tilde{3}.
\label{colourframon}
\end{equation}
\end{itemize}
Here we have left out some spacetime independent factors which do not enter 
into the calculations in this paper.  

The flavour framon $\bphi$ (written as a vector here) transforms as a doublet in local $su(2)$, 
while the colour framon $\bPhi$ (written as a matrix here)
transforms as a triplet in local $su(3)$ but as an anti-triplet in global 
$\widetilde{su}(3)$.  

The colour framon, being coloured and colour being confining, cannot propagate 
as a particle in free space, but can combine with a colour anti-triplet 
to form a colour singlet particle which propagates.  In
particular, it can combine with an antiframon to give freely propagating bosons.
Here, we are interested only in the $p$-wave bound states: $\bPhi^\dagger D_\mu 
\bPhi$, which we call $G$s.  There are 8 of them, conveniently labelled by the
Gell-Mann matrices $\lambda_1, \ldots, \lambda_8$.  They are the colour analogues
of the $W$-bosons labelled by the Pauli matrices $\tau_1, \tau_2, \tau_3$ in the 
electroweak sector.  The analogy  is not so clear in the conventional picture where
the electroweak symmetry $su(2)$ is considered as spontaneously broken.  But in
the confinement picture of 't~Hooft, which he showed in an illuminating paper
\cite{thooft} to be a ``mathematically equivalent'' interpretation, the $W$
appear also as $\bphi^\dagger D_\mu \bphi$ bound states, but here via $su(2)$
flavour confinement.  Now in the electroweak theory, $W^3_\mu$ mixes with the
$u(1)$ gauge field $A_\mu$ to form $\gamma$ and $Z$.  So, in parallel, one 
would not be surprised that in the FSM when the symmetry is extended, the
mixing is extended to include also $G^8$.

Explicitly how this mixing goes will be worked out in the next section.  We
start here with some general observations.  The mixing of $A_\mu$ and $W^3_\mu$
in the electroweak theory involves three paramenters which we may take as $g_1$,
$g_2$ and $\zeta_W$, the last being the vacuum expectation value  of the standard Higgs field, which 
is essentially the same as the flavour framon $\bphi$ above.  When extended to 
include $G^8$ in the FSM, the mixing involves two more parameters: $g_3$ and 
$\zeta_S$, the former being the colour coupling as measured in several processes \cite{pdg},
the latter the vacuum value of the colour framon $\bPhi$.  The fit to data 
performed in \cite{tfsm} suggests that $\zeta_S$ has a value of order TeV, 
leaving the new extended mixing scheme with rather little freedom.

As we shall see, the parameters $\zeta_W$ and $\zeta_S$ appear in the extended
mixing formulae in the combination $\zeta_W^2/\zeta_S^2$ which is of order 
$10^{-2}$.  If this should lead to deviations from the standard mixing formulae
by amounts of that order, then it would be a disaster, for the standard mixing
scheme has already been checked by experiment to an  accuracy of  several orders higher.
This explains why tests of the new mixing scheme are of particular
acuteness and urgency.  Fortunately, as will be shown, in the predictions of
the new mixing scheme of the mass and decay widths of the $Z$, the quantities so 
far studied, there are subtle cancellations which mean that
the deviations from the SM are of 
order $\zeta_W^2/\zeta_S^2$ multiplied by the small  factor $g_1^4$,
which for $\zeta_S = 2$ TeV, turn out to be of order  $\sim 10^{-4}$
and within the present experimental limits.  In other words, for $\zeta_S \geq$
2 TeV, the FSM has survived so far the tests against experiment to which it has
been subjected.

Turning the argument around and taking the positive view that these deviations
of FSM from SM are new physics to be searched for, we note that they depend on 
only one parameter of uncertain value, $\zeta_S$, and are thus all correlated.  
Besides, they are also correlated to the $G$ mass (which is
proportional to $\zeta_S$ to a very good approximation), 
while $G$ itself can appear in LHC experiments as a lepton-antilepton
enhancement in the multi-TeV range.  The whole complex would thus appear to be 
a fruitful region for future experiments to explore.

\section{Mixing in the $\gamma-Z-G$ Complex}

The mass (squared) matrices of the vector bosons can be extracted from the kinetic 
energy terms of the framons in the Lagrangian \cite{cfsm}, namely:
\begin{eqnarray}
& \!\!\!\!\! [(D_\mu \bphi)^\dagger D_\mu \bphi], & D_\mu = \partial_\mu + \half i g_1
   A_\mu -\half ig_2 B_\mu,   \nonumber \\
&\Tr[(D_\mu \bPhi)^\dagger D_\mu \bPhi], & D_\mu = \partial_\mu -
                                           i g_1  \Gamma A_\mu -\half  ig_3 C_\mu, 
\end{eqnarray} 
where the flavour framon $\bphi$ has  charge $-\half$, $A_\mu, B_\mu, 
C_\mu$ are respectively the gauge potentials of the gauge symmetries $u(1),su(2),
su(3)$,  and $\Gamma$ is the charge matrix of the colour framon:\footnote{It 
is understood that, in our convention where in the matrix $\bPhi$, rows are 
labelled by local colour $a$ but columns by global colour $\tilde{a}$, the matrix 
$\Gamma$ is to operate on $\bPhi$ from the right.}
\begin{equation} 
\Gamma = \left( \begin{array}{ccc} -\third & 0 & 0 \\
                                     0 & -\third & 0 \\
                                     0 & 0 & +\twothirds \end{array}
                                 \right), 
\end{equation}
in which the charges have been chosen explicitly to keep the photon massless.

First consider the flavour framons. Adopting the confinement picture of 't~Hooft
(which is more immediately extendable later to the colour framons) we can use 
the gauge freedom to fix the gauge by rotating, with an $su(2)$ transformation 
$\Omega(x)$, the  doublet scalar field $\bphi$ to point, at every 
spacetime point $x$, in the first direction and to be real, 
\begin{equation}
\bphi = \Omega \left( \begin{array}{c}  \rho \\0 \end{array} \right)
       = \Omega \bphi_{GF},
\end{equation}
with $\rho$ real.  We can thus write:
\begin{equation}
\rho = \zeta_W + h_W,
\label{rho}
\end{equation}
where, $\zeta_W$ is the vacuum expectation value of $\rho$, and $h_W$, as its fluctuation about the 
vacuum value, is the Higgs boson field \cite{thooft}, \cite{banksrab}.

Using the gauge-fixed field $\bphi_{\rm GF}$ and introducing the gauge
invariant quantity:
\begin{equation}
\half \tilde{B}_\mu = \tfrac{i}{g_2} \Omega^\dagger (\partial_\mu -
\half i g_2 B_\mu) \Omega,
\label{Btilde}
\end{equation}
so that 
\begin{equation}
\Omega^\dagger D_\mu \Omega = \half i g_1 A_\mu - \half i g_2
\tilde{B}_\mu,
\end{equation}
we can rewrite the kinetic energy term to leading order (since
$\bphi_{\rm GF}$ is constant to that order),
\begin{equation}
[D_\mu \bphi]^\dagger [D_\mu \bphi] \, = \,  \bphi_{GF}^\dagger [+ig_1 \half A_\mu 
  -ig_2\tilde{B}_\mu]^\dagger \, \, [ +ig_1 \half A_\mu - ig_2 \tilde{B}_\mu] \bphi_{GF}.
\end{equation}
This gives for the mass term, $\tilde{B}_\mu$ being hermitian:
\begin{equation}
(\zeta_W, 0) \quart \left[ g_1^2 A_\mu^2 + g_2^2 \tilde{B}_\mu^2
   - 2g_1 g_2 A_\mu \tilde{B}_\mu \right] \left( \begin{array}{c} \zeta_W \\0 
   \end{array} \right),
\end{equation}  
that is,  $\zeta_W^2$ times the 11 element of the quantity inside the
square bracket.  We thus obtain the following mass squared matrix 
\begin{equation}
\frac{\zeta_W^2}{4} \left( \begin{array}{cccc} g_2^2 & 0 & 0 & 0 \\
                           0 & g_2^2 & 0 & 0 \\
                           0 & 0 & g_2^2 & -g_1 g_2 \\
                           0 & 0 & -g_1 g_2 & g_1^2 
                           \end{array} \right),
\label{massmatew}
\end{equation}
with only the $\tilde{B}^3$ component mixing with the $A_\mu$, giving
the massless photon and the SM $Z$ boson:
\begin{eqnarray}
\gamma_\mu &=& \frac{1}{\sqrt{g_1^2 + g_2^2}} \left( g_2 A_\mu + g_1
               \tilde{B}^3_\mu \right) \nonumber \\
Z_\mu &=& \frac{1}{\sqrt{g_1^2 + g_2^2}} 
   \left( -g_1 A_\mu + g_2 \tilde{B}^3_\mu \right),
\label{Zmu}
\end{eqnarray}
with
\begin{equation}
\frac{1}{e^2} = \frac{1}{g_1^2} + \frac{1}{g_2^2}.
\label{SMcouplings}
\end{equation}

Next we consider the kinetic energy term of the colour framons (\ref{colourframon}).
Again we can use an $su(3)$ matrix $\Omega(x)$ to fix the gauge of the
framons $\bPhi$, 
\begin{equation}
\bPhi = \Omega \bPhi_{\rm GF},
\end{equation}
though not completely, there not being sufficient degrees of freedom in $su(3)$ 
to do so, but still enough to make its vacuum value take on a 
diagonal form as follows:
\begin{equation}
\bPhi_{\rm GF} \rightarrow \frac{\zeta_S}{\sqrt{3}} \left( \begin{array}{ccc}
         \sqrt{1 - R} & 0 & 0 \\
         0 & \sqrt{1 - R} & 0 \\
         0 & 0 & \sqrt{1 + 2R} \end{array} \right) = \bPhi_{\rm vac},
\end{equation}
and it is this last form which gives us the mass matrix of the vector 
bosons.  Here $R$ is a scale-dependent quantity made out of coupling
constants of the terms in the FSM framon potential, and its actual
value at any scale has been obtained by fitting FSM to fermion masses
and mixing data \cite{tfsm}.  For $\zeta_S \geq 2$ TeV which interests us here,
one can effectively put $R = 0$.

Proceeding as in the flavour case, we define
\begin{equation}
\half \tilde{C}_\mu = \tfrac{i}{g_3} \Omega^\dagger (\partial_\mu -\half ig_3 C_\mu) \Omega.
\end{equation}
so that the kinetic energy term can be rewritten to leading order as
\begin{eqnarray}
&& \Tr \left[ \bPhi_{GF}^\dagger \Omega^\dagger D_\mu^\dagger \Omega \Omega^\dagger
            D_\mu \Omega \bPhi_{GF} \right]
     \nonumber \\
&& = \, \Tr \left[\bPhi_{GF}^\dagger \left(- ig_1 \Gamma A_\mu -\half ig_3 \tilde{C}_\mu \right)^\dagger 
       (- ig_1 A_\mu \Gamma -\half ig_3 \tilde{C}_\mu) \bPhi_{GF} \right],
\label{KECt}
\end{eqnarray}
We now expand $\tilde{C_\mu}$ in terms of the usual Gell-Mann
matrices:
\begin{equation}
\tilde{C_\mu}= \sum_K \tilde{C_\mu}^K \lambda_K
\end{equation}
and rewrite the kinetic energy term as
\begin{eqnarray}
{\rm K.E.} &=& \Tr \Bigg[ 
\bPhi_{\rm vac}^\dagger\bPhi_{\rm vac} \, \Bigg( g_1^2 A_\mu A^\mu \Gamma \Gamma^\dagger \nonumber \,+
\half g_1g_3 A^\mu \Gamma \left( \sum \tilde{C_\mu}^K \lambda_K \right) \,
\nonumber \\
  & + & \half g_1g_3 A_\mu  \left( \sum \tilde{C_\mu}^K\lambda_K \right) \Gamma^\dagger
\, + \, \quart g_3^2 \Big( \sum_K \tilde{C_\mu}^K\lambda_K\Big)
\Big( \sum_L \tilde{C_\mu}^L \lambda_L\Big) \Bigg)\, \Bigg].
\end{eqnarray}

Now because we have chosen 
\begin{equation}
\Gamma = -\frac{1}{\sqrt{3}} \lambda_8
\end{equation}
the resulting mass matrix is block-diagonal, with only the $0$-$8$ block
non-diagonal:
\begin{equation}
\tfrac{2}{3} (1+R) \zeta_S^2 \left( \begin{array}{cc} \tfrac{1}{3}
                                      g_1^2 & -\tfrac{1}{\sqrt{3}}
                                              g_1g_3 \\
-\tfrac{1}{\sqrt{3}}g_1g_3 & g_3^2 \end{array} \right).
\end{equation}
This has a zero mode, as expected.

Finally we consider the sum of the two kinetic energy terms, and
concentrate on the ($ 3 \times 3$) $A-\tilde{B}^3-\tilde{C^8}$ mixing matrix as (where
the order of the rows and columns are labelled according to the
symmetries $u(1), su(2), su(3)$)
\begin{equation}
M= \left( \begin{array}{ccc}
(\ell+\third k)\,g_1^2 & 
-\ell g_1 g_2 & -\frac{k}{2\sqrt{3}}\,g_1 g_3 \\
\vspace*{1mm} \\
-\ell g_1 g_2 & \ell g_2^2 & 0\\
\vspace*{1mm} \\
 -\,\frac{k}{2\sqrt{3}} g_1 g_3 & 0 & \frac{k}{4}\,g_3^2 
\end{array} \right) 
\label{mamatrix}
\end{equation}
where
\begin{equation}
\ell = \quart \zeta_W^2, \ \ k=\tfrac{2}{3} (1+R) \zeta_S^2
\label{kl}
\end{equation}
Crucially this has again a zero mode, which is indispensable for a
viable theory containing the photon.

Apart from the zero mode, the other two eigenvalues ($\lambda_\pm$) are given by the
roots of the following quadratic equation
\begin{equation}
\lambda^2 \, - \, \lambda \left( \ell g_2^2 \, + \, \tfrac{k}{4} g_3^2 +(\ell +                  
\tfrac{k}{3}) g_1^2 \right) \, + \, k \ell \left(\quart g_1^2 g_3^2 + \quart g_2^2                
g_3^2 + \third g_1^2g_2^2 \right) \, = \,0. 
\label{quadratic}
\end{equation}
These are then the mass squared of respectively the $Z$ boson and
another boson we call $G$ of higher mass not present in the SM spectrum.  
The discriminant of (\ref{quadratic}) is positive for all values of the coupling
constants, as it can be re-written as:
\begin{equation}
\left( \ell g_2^2 +(\ell-\tfrac{k}{3}) g_1^2 -\tfrac{k}{4} g_3^2 \right)^2 +
\tfrac{4}{3} k \ell g_1^4.
\label{discrimpos}
\end{equation}
In this form, one sees immediately that both roots are positive
for all values of the coupling constants.

The normalized zero eigenvector (which corresponds to the photon) is given by
\begin{equation}
 v_1= \left( \begin{array}{c} \frac{e}{g_1} \\ \vspace*{0.3mm}\\
\frac{e}{g_2} \\\vspace*{0.3mm}\\  \frac{2}{\sqrt{3}}                          
            \frac{e}{g_3} \end{array} \right) 
\label{photonvec}
\end{equation}
where
\begin{equation}
 \frac{1}{e^2} = \frac{1}{g_1^2} + \frac{1}{g_2^2} +
  \frac{1}{\tfrac{3}{4} g_3^2}.
\label{FSMcouplings}
\end{equation}
That this is indeed the correct normalization can be checked by
writing the $su(2) \times u(1)$ neutral current in terms of the mass
eigenstates and identifying the electromagnetic current as the piece that is
coupled to the photon, the coefficient of which will give the electric
charge $e$. 

Comparison with equation (\ref{SMcouplings}) immediately tells us that
the coupling $g_1$ in FSM differs from that of the SM.  The exact relation
is given in equation (\ref{relation}) below.

Then, after some more algebra, the other two eigenvectors can be found and one 
obtains the mixing matrix:
\begin{equation}
 \left( \begin{array}{c} \gamma_\mu \\ \vspace{2pt}\\Z_\mu \\ \vspace{2pt}\\
G_\mu  \end{array} \right) =
\left( \begin{array}{ccc}
            \tfrac{e}{g_1} & \tfrac{e}{g_2} & \tfrac{2}{\sqrt{3}}
                                             \tfrac{e}{g_3} \\
\vspace{2pt}\\
-X_- & Y_- & -W_- \\ \vspace{2pt}\\ X_+ & -Y_+ & W_+ \end{array} \right)
\left( \begin{array}{c} A_\mu \\ \vspace{2pt}\\ \tilde{B}_\mu^3  \\ 
\vspace{2pt}\\ 
\tilde{C^8_\mu} \end{array} \right),
\label{mixing}
\end{equation}
where 
\begin{eqnarray}
X_\pm &=& \left( kA-3N_2^2 \lambda_\pm \right) \frac{g_1}{N_2 N_\pm} - \frac{k g_1^2
          g_2 \sqrt{A}}{N_3 N_\pm}
\nonumber \\
Y_\pm &=& \left( kA-3N_2 ^2 \lambda_\pm \right) \frac{g_2}{N_2 N_\pm} + \frac{k g_1^3
         \sqrt{A}}{N_3 N_\pm}\nonumber \\
W_\pm &=& \frac{\sqrt{3}}{2} \left( kg_1^2 \sqrt{A} \right) \left( \frac{g_1^2
          +g_2^2}{g_1g_2} \right) \frac{g_3}{N_3 N_\pm}
\label{X-Y-}
\end{eqnarray}
And the normalization factors are
\begin{eqnarray*}
N_2^2 &=& g_1^2 + g_2^2 \\
N_3^2 &=& \frac{g_1^2 + g_2^2}{g_1^2 g_2^2} A \\
N_\pm^2 &=& (kA - 3 N_2^2 \lambda_\pm )^2 + k^2 g_1^4 A
\end{eqnarray*}
with
\begin{equation}
A= g_1^2g_2^2 +  \tfrac{3}{4} g_1^2 g_3^2 +  \tfrac{3}{4} g_2^2 g_3^2.
\end{equation}

These mixing formulae of the FSM differ considerably in form from those of the 
SM.  On the other hand, the mixing scheme of the SM has been tested already to
great accuracy by experiment over a wide range of phenomena.  Hence, for the
FSM to remain viable, it has to be hoped that, despite the difference in form,
the deviations in predicted values of the measured quantities would still 
remain within the present experimental errors over the whole range of data on
which the SM mixing has been tested.  However, to check if this is true will 
be a long and arduous process requiring a similar degree of sophistication to 
that applied in checking the standard model, which is not yet available in the 
FSM where the necessary tools have not been developed.
Nevertheless, we believe that it is imperative to do such checks to
the extent they can be done in the FSM at present before we attempt to
go further with the programme, and 
we shall do so here for
the most urgent tests which can immediately be 
implemented.

In view of the present limitations of the FSM and our specific aim of testing
just the change in mixing from the SM to the FSM, we have devised the following 
procedure specially tailored for the purpose:
\begin{itemize}
\item {\bf (A)}
At present, one is not in a position to calculate loop corrections in general
in the FSM, not having yet developed the tools for doing so.  However, to check 
with the data at present accuracy, tree-level results are not adequate.  
We propose therefore to adopt the following as a test criterion.  It is widely 
accepted, and we will thus take for granted, that the loop-corrected predictions 
of the SM are in agreement with experiment within present errors.  Then, 
assuming that the difference in loop corrections between the SM and FSM is of 
higher order in smallness, we consider  that the difference between their 
tree-level results  should be a good estimate already of 
the loop-corrected difference.  Hence, in what follows, if it is found that the 
tree-level prediction of FSM for a certain quantity deviates from the 
tree-level prediction of the SM  by less than the present 
experimental error, we would consider that the FSM has passed the test, 
regardless of whether the tree-level predictions themselves of either the SM 
or the FSM are within errors of the experimentally measured values.
\item{\bf (B)}
At tree level, the $W^\pm$ bosons are not affected by the change in
mixing of  the neutral partners of the FSM from the SM.
We may thus choose to regard the $W$ mass and width as given and 
determine from them the two parameters $g_2$ and $\zeta_W$.    Then 
together with $e$ for the photon, 
and the colour coupling $g_3$, independently measured from $Z$ decays and perturbative QCD, we have all the parameters needed for the above mixing 
schemes (both of the SM and the FSM), except for $\zeta_S$ for which there is 
only a very crude order of magnitude estimate \cite{tfsm,cfsm}.  We may thus
choose to formulate our test of the FSM mixing scheme as follows.  We calculate 
with these parameters the tree-level properties of the $Z$ as predicted by respectively 
the SM and FSM mixing schemes, and compare the results.  And if the two answers 
for each quantity differ by less than the experimental error, then by {\bf (A)} 
above, we shall consider that the FSM has passed the test.  This may not be the 
usual procedure taken, which tends more to start with the $Z$, for which errors 
are generally smaller, to predict the $W$.  But since we fix the
relevant parameters from the $W$ data, this makes the logic clearer,
and is equivalent.
\end{itemize}

Taking then from the PDG tables \cite{pdg}
\begin{eqnarray}
\zeta_W &=& 246\ {\rm GeV} \nonumber \\
m_W &=& 80.385\ {\rm GeV},
\label{Wparam}
\end{eqnarray}
and determining $g_2$, one has then, together with $e$ and $g_3$ from the PDG
tables and $\zeta_S$ from \cite{cfsm}, the following list of parameters taken at the scale of the $Z$ mass:
\begin{eqnarray}
e^2 &=& 0.098175 (= 4\pi/128) \nonumber \\
g_2^2 &=& 0.4271 \nonumber \\
g_3^2 &=& 1.4828 \nonumber \\
\zeta_S & \sim & {\rm order\ TeV}.
\label{fixparam}
\end{eqnarray}  
We recall that the values
of $g_1$ differ in the two mixing schemes.  We shall reserve the symbol
$g_1$ for FSM and denote the SM value as $g_1^{\rm SM}$.
\begin{eqnarray}
g_1^2 &=& 0.1440  \nonumber \\
(g_1^{\rm SM})^2 &=&0.1275  \nonumber \\  
\label{addparam}
\end{eqnarray}

With these parameters, we shall calculate:
\begin{itemize}
\item the $Z$ mass,
\item the decay widths of $Z$ into $\ell^+ \ell^-$ and $q \bar{q}$ pairs.
\end{itemize}
and show that provided one chooses $\zeta_S \geq 2$ TeV, the difference
between the two mixing schemes is within the present experimental errors.
By {\bf (A)} then, we conclude that the FSM has so far passed the test.

We note that the values quoted in (\ref{Wparam}), (\ref{fixparam}) and
(\ref{addparam}) are all just central values each with an error which has not
been displayed.  Thus all predictions deduced from them on the $Z$ mass and 
decay widths will inherit an error from them and these errors will have to 
be accounted for when applying criterion {\bf (A)}.  This is particularly 
relevant here when deducing $Z$ properties from the $W$ since the experimental 
errors on the $W$ are generally bigger than on the $Z$.  We shall leave for 
later the details how this point will be accounted for in each case.

\section{The $Z$ mass}

\subsection{Exact and approximate formulae}

The tree-level mass of the Z is given by the smaller of the two roots of
(\ref{quadratic}):
\begin{eqnarray}
m_Z^2&=&
\frac{1}{2} \left( (\ell g_2^2 + \tfrac{k}{4} g_3^2 +(\ell +
  \tfrac{k}{3}) g_1^2) \right.  \nonumber \\
 &&- \left. \sqrt{((\ell g_2^2 + \tfrac{k}{4} g_3^2 +(\ell +
  \tfrac{k}{3}) g_1^2)^2 - k \ell (g_1^2 g_3^2 +g_2^2                
  g_3^2 +\tfrac{4}{3} g_1^2g_2^2)} \right)
\label{zmass}
\end{eqnarray}

For comparison later with the SM, we expand this in powers of $(\ell/k)$, which
according to (\ref{fixparam}) is a small parameter.  To zeroth order, one has:  
\begin{equation} 
m_Z'^2  = \frac{\tfrac{3}{4} g_1^2 g_3^2 + g_1^2 g_2^2 + \tfrac{3}{4}g_2^2 g_3^2}
{g_1^2 + \tfrac{3}{4} g_3^2}\,\, \ell \, 
\, =\, \, \frac{\ell A}{B} 
\label{zmassappr}
\end{equation}
This means first that, provided $\zeta_W^2/\zeta_S^2$ is small, the FSM mass is 
a good approximation of the SM mass, both evaluated at tree level.
Furthermore,
to this order, $m_Z$ is independent of $k$  so that we do 
not need to know the value of $\zeta_S$. In the last equality of (\ref{zmassappr})
$A$ was defined above and $B=g_1^2 + \tfrac{3}{4} g_3^2$, which will be useful 
later. 

If we expand further, we get

\begin{equation}
m_Z^2= \ell \left( g_1^2 - \third \tfrac{g_1^4}{\quart g_3^2 + \third
g_1^2} - \tfrac{\ell}{k} \tfrac{1}{9} \tfrac{g_1^4}{(\quart g_3^2 + \third
g_1^2)^3} A \right) + \cdots ,
\label{mzfirst}
\end{equation}
which can be rewritten using (\ref{zmassappr}) as
\begin{equation}
m_Z= m_Z^{\rm SM} \left( 1 -  g_1^4 \tfrac{\ell}{k} \tfrac{1}{18}
\tfrac{B}{(\quart g_3^2 + \third g_1^2)^3}  +\cdots \right).
\label{mzfirstbis}
\end{equation}
We see that, for our benchmark value of $\zeta_S \sim 2$ TeV,
the first order $\ell/k$ (itself of order $10^{-2}$) term is
multiplied by 
$g_1^4$ (of similar order), making the whole term of order $10^{-4}$.   

Furthermore, equation (\ref{mzfirstbis}) shows that the
difference between the FSM and SM values for our benchmark  $\zeta_S
\sim 2$ TeV is of order $10^{-4}$, which would bring it to about the present
experimental error.  That this is indeed the case will be shown explicitly in
what follows.

\subsection{Comparison with SM}

From the relations (\ref{SMcouplings}) and (\ref{FSMcouplings}) for respectively
the SM and FSM, we deduce the relation:
 \begin{equation}
\frac{1}{\gsm^2} =  \frac{1}{g_1^2} + \frac{1}{\tfrac{3}{4} g_3^2} =
\frac{g_1^2 +\tfrac{3}{4} g_3^2}{\tfrac{3}{4} g_1^2 g_3^2}.
\label{relation}
\end{equation}
Using (\ref{relation}) we can then easily work out that
\begin{equation}
(m'_Z)^2 = \ell (\gsm^2 +g_2^2) = (m_Z^{\rm SM})^2.
\label{mZ}
\end{equation}

This means first that, provided $\zeta_W^2/\zeta_S^2$ is small, the FSM mass is 
a good approximation of the SM mass, both evaluated at tree level.  Secondly,
because the first order $\ell/k$ term vanishes, the difference between the FSM 
and SM values is {\it not} of order $\ell/k \sim 10^{-2}$ which would have been 
disastrous, but of order $(\ell/k)^2 \sim 10^{-4}$, for our benchmark value of
$\zeta_S \sim 2$ TeV, which would bring it to about the present experimental 
error.  That this is indeed the case will be shown explicitly in what follows.

\subsection{Numerical results and comparison with experiment}

We recall that before any mixing scheme is implemented, whether in the SM or
the FSM, the $Z$ and the $W$ are degenerate in mass.  What mixing does is to
introduce a shift in the masses $m_{\rm shift} =m_Z-m_W$, and it is
precisely
this quantity the models predict which should be compared to experiment.

The PDG \cite{pdg} gives
\begin{equation}
m_{\rm shift} =m_Z-m_W = 10.803 \pm 0.015\ {\rm GeV},
\label{shiftexp}
\end{equation}
with an impressively small error:
\begin{equation}
\Delta m_{\rm shift}^{\rm exp} = 15\  {\rm MeV}.
\label{expDelta}
\end{equation}

According to criterion {\bf (A)}, this is to be compared to the difference
in
mass shifts obtained respectively in the SM and FSM: $m_{\rm shift}^{\rm SM}
- m_{\rm shift} = \Delta m_{\rm shift}$.  Now $m_W$ being the same in SM and
FSM, we have
\begin{equation}
\Delta m_{\rm shift} = \Delta m_Z = m_Z^{\rm SM} - m_Z.
\label{shift-delta}
\end{equation}

Using the values in (\ref{fixparam}) and (\ref{addparam}), and the
expressions
(\ref{zmass}), and (\ref{zmassappr}), we obtain for our benchmark value of 2
TeV for $\zeta_S$:
\begin{eqnarray}
m_Z &=&   91.5884\ {\rm GeV} \nonumber \\
m_Z^{\rm SM} &=& 91.5988\ {\rm GeV},
\label{zmasscompared}
\end{eqnarray}
giving
\begin{equation}
\Delta m_{\rm shift} = \Delta m_Z = 10.4\ {\rm MeV}.
\label{Deltashift}
\end{equation}
We thus indeed have:
\begin{equation}
\Delta m_{\rm shift} = 10.4\ {\rm MeV}
   < 15\ {\rm MeV} = \Delta m_{\rm shift}^{\rm exp},
\label{compDelta}
\end{equation}
as criteriom {\bf (A)} requires for the FSM to pass the test.

Notice that both the tree-level predictions for the $Z$ mass
of the two models (\ref{zmasscompared}) lie outside present
experimental bounds
\begin{equation}
m_Z^{\rm exp}=91.1876 \pm 0.0021\ {\rm GeV},
\end{equation}
which shows the inadequacy of tree-level approximations at
present experimental accuracy, and hence the relevance of
criterion {\bf (A)} for testing the FSM at tree level.  Note
also that in comparing $m_{\rm shift}$ to experiment instead
of $m_Z$ directly, we have folded in the error on $m_W$ which
at present is larger than the error on $m_Z$ itself.

It will be shown in the last section that the difference
between the predicted values decreases with increasing
$\zeta_S$ so that criterion {\bf (A)} remains satisfied for
all $\zeta_S$ above 2 TeV.

\section{$Z$ decay into a fermion-antifermion pair}

\subsection{Exact and approximate formulae}

From equation (\ref{mixing}) we can write
\begin{equation}
Z_\mu= -X_- A_\mu + Y_-\tilde{B}_\mu^3 - W_- \tilde{C}_\mu^8.
\label{zstate}
\end{equation}
The last term does not contribute since $\tilde{C}_\mu$ does not couple directly
to quarks and leptons, and we shall ignore it henceforth.  The $su(2)$ part of 
the $Z$ couples to only the left-handed fermion, while the $u(1)$ part of $Z$ 
couples to both left and right handed fermions, though differently.  

We can now write the current for $Z$ coupled to the fermion field $f$:
\begin{eqnarray}
\!\!\!\!\!\!\!\!\!\!\!j_\mu\!\!\! &=&\!\!\! -\half \, g_1 \, X_- \,
\left[ \left(\, \bar{f}_L \, \left( 2Q - 2I_3 \right) \gamma_\mu \, f_L \, \right) \, + \, \left(\, \bar{f}_R \, \left( 2Q \right) \gamma_\mu\, f_R \,\right) \right] 
 \, + \, g_2 \, Y_- \, \left(\, \bar{f}_L\, I_3 \, \gamma_\mu \, f_L \,\right) \nonumber \\
&=&
\!\!\!\bar{f}\,\gamma_\mu\, \left[-g_1 X_-\left(Q-I_3 \right) \,+ \, g_2 \, Y_- \,I_3\, \right] \,\frac{1-\gamma_5}{2}\,\,f  \, - \, \bar{f} \,
\left[ \,g_1   X_-Q \,\right] \,\frac{1+\gamma_5}{2} \,f \nonumber \\
&=&
\!\!\!\half \, \bar{f} \,\gamma_\mu\,\Big[ \, -2 g_1 X_- Q \, + \left( \,g_1 X_- \, + \,g_2 Y_- \right) I_3 
\,-\, \left(\,g_1 X_- + g_2 Y_- \,\right) I_3 \,\gamma_5 \,\Big]\,f.\nonumber \\
\label{current}
\end{eqnarray}

We define the vector and axial vector couplings $c_V$ and $c_A$ in the standard form:
\begin{eqnarray}
c_V &=& g_1X_-(-2Q+  I_3) + g_2Y_-I_3 \nonumber \\
c_A &=&g_1 X_- I_3 + g_2 Y_- I_3
\label{caandcv}
\end{eqnarray}
so that
\begin{equation}
j_\mu = \half \bar{f}\gamma_\mu\,(c_V -c_A \gamma_5)\,f.
\end{equation}

Hence the $Z$ vertex factor is $ -i \half \gamma_\mu (c_V-c_A \gamma_5)$, from 
which using standard formulae \cite{thomson}, in the case where the fermion masses can be 
neglected, we can write the partial width as:
\begin{equation}
\Gamma (Z \to f \bar{f}) = \frac{m_Z}{48 \pi} (c_V^2+c_A^2). 
\label{gamma}
\end{equation}

This formula is often re-expressed in terms of the very accurately
determined Fermi constant $G_F$, with an accuracy of 500 ppb \cite{pdg}:
\begin{equation}
\Gamma (Z \to f \bar{f}) = \frac{G_F m_Z^3}{6 \sqrt{2} \pi}
\frac{(c_V^2+c_A^2)}{g_Z^2},\ \ G_F=1.166378 \times 10^{-5}\; {\rm GeV}^{-2},
\label{fermi}
\end{equation}
where $g_Z^2= \gsm^2 + g_2^2=A/B. $
This is the formula we shall use.

From this, again for later comparison with the SM, we next derive approximate 
formulae, by expanding in powers of $(\ell/k)$.  First, dropping all terms 
depending on $(\ell/k)$ we obtain from equation (\ref{X-Y-}) the zeroth order
approximations:
\begin{eqnarray*}
X_-' &=& \frac{kA g_1}{N_2N_-} - \frac{k g_1^2 g_2 \sqrt{A} }{N_3 N_-}
= \frac{kg_1}{N_2 N_-} (A-g_1^2 g_2^2) \\
Y_-'&=& \frac{kA g_2}{N_2 N_-} + \frac{kg_1^3 \sqrt{A}}{N_3 N_-}
= \frac{k g_2}{N_2N_-} (A+g_1^4)\\
N_-'^2 &=& k^2 A(A+g_1^4)
= k^2 A B (g_1^2 + g_2^2),
\end{eqnarray*} 
giving 
\begin{equation}
X_-'= \frac{g_1}{\sqrt{AB}} \left( \frac{3}{4} g_3^2 \right), \quad Y_-' = g_2
\sqrt{B/A}.
\end{equation}
Substituting into (\ref{caandcv}), we obtain
\begin{eqnarray}
c_V' &=& \frac{-\tfrac{3}{2} g_1^2g_3^2 Q + A I_3}{\sqrt{AB}} \nonumber
  \\
c_A' &=& \sqrt{\frac{A}{B}} I_3
\label{cvca-approx}
\end{eqnarray}
Hence the zeroth order partial width is
\begin{equation}
\Gamma' (Z \to f \bar{f}) = \frac{G_F m_Z^3}{6 \sqrt{2} \pi}
\frac{(c_V'^2+c_A'^2)}{g_Z^2} .
\label{gammaapprox}
\end{equation}

Next we look at order $\ell/k$, and find that 
\begin{equation}
X_-= \frac{g_1}{\sqrt{AB}} \left(\frac{3}{4} g_3^2 -\frac{3A}{B^2} g_1^2
\left(\frac{\ell}{k}\right)\right) + {\cal O} \left(\frac{\ell^2}{k^2} \right),
\end{equation}
which has an $\ell/k$ term multiplied by $g_1^3$, while $Y_- $ has no $\ell/k$ 
term.   Now in both $c_V$ and $c_A$, we have $X_-$ multiplied again by $g_1$ 
so that the first order term in $\ell/k$ in the decay width is in fact 
multiplied by $g_1^4$ which according to (\ref{addparam}) 
is approxmately  $0.02$.  As in the $Z$ mass case, this fact is of 
significance when comparing with the SM and with experiment. 

\subsection{Comparison with SM}

The expression in SM for the partial width into a fermion pair is given
by:
\begin{equation}
\Gamma^{\rm SM} (Z \to f \bar{f}) = \frac{g_Z^2 m_Z}{48 \pi} ((c_V^{\rm SM})^2 +
 (c_V^{\rm SM})^2),
\label{pw-sm}
\end{equation}
and the vector and axial vector couplings 
in SM are given by
\begin{eqnarray}
c_V^{\rm SM} &=& -2Q \sin^2 \theta_W + I_3 \nonumber \\
c_A^{\rm SM} &=& I_3
\label{smcvca}
\end{eqnarray}

Substituting in $\sin \theta_W =e/g_2$ we can easily work out that, to zeroth 
order in $(\ell/k)$,
\begin{eqnarray}
c_V \rightarrow c'_V &=& g_Z  c_V^{\rm  SM} \\
c_A \rightarrow c'_A &=& g_Z  c_A^{\rm  SM},
\end{eqnarray}
which means that, comparing with equation (\ref{gamma}),
\begin{equation}
\Gamma' (Z \to f \bar{f})  = \Gamma^{\rm SM} (Z \to f \bar{f}) .
\end{equation}

So once again, as is the case for $m_Z$, the $Z$ decay widths in the FSM in
the limit $\ell/k \sim \zeta_W^2/\zeta_S ^2 \to 0$ is identical to the
SM prediction.
We may ask why it should be so, since in the FSM the $Z$ has a component in 
$su(3)$, which introduces two new parameters into the problem, namely, besides 
the strong vacuum expectation $\zeta_S$, here effectively put to $\infty$, also 
the colour coupling $g_3$.  There is perhaps a deep reason for this that
we have not yet fathomed, but we can observe that in $m'_Z$ and $\Gamma'$ the 
couplings $g_1$ and $g_3$ always come in the combination (\ref{relation}) so 
that the two together can be replaced by $g_1^{\rm SM}$.  This does not in 
itself prove that the expressions must be equal, but it does at least provide 
an indication how it can occur.


Recalling next from 4.1 that the $\ell/k$ term in the decay width $\Gamma$ is 
multiplied by $g_1^4 \sim 0.02$, we conclude that the FSM prediction for the
width will differ from that of the SM only by order $g_1^4 (\ell/k) \sim 
10^{-4}$, that is,  similar to the deviation in the predictions for the $Z$ mass.
This will explain why numerically the deviations of the FSM from the SM are found 
to remain within the stringent experimental bounds, as shown below.   
     

\subsection{Numerical results and comparison with experiment}

Using the parameter values (\ref{fixparam}) and (\ref{addparam}) with the
benchmark value $\zeta_S = 2$ TeV, we obtain the mixing matrix
(\ref{mixing})
as:
\begin{equation}
\left( \begin{array}{lll}
0.8257 &    0.4794 &      0.2971\\
  -0.4507  &   0.8776  &    -0.1635 \\
 -0.3392  &  0.0011  &  0.9407
\end{array}
\right).
\label{mixingnum}
\end{equation}

Using the numerical values of $X_-$ and $Y_-$ above, we can evaluate
the different partial widths into fermion pairs\footnote{The formulae
(\ref{gamma}),  (\ref{fermi}),
(\ref{gammaapprox}) and (\ref{pw-sm}) hold in the limit of zero fermion
masses, which should be a good enough aproximation for present purposes.
Hence they are identical for all charged leptons ($e, \mu, \tau$),
identical for the $U$-type quarks ($u,c$), and also identical for all
$D$-type quarks ($d,s,b$).  For definiteness, we quote for comparison
only the experimental results for the decays $Z \to e^+e^-$, and $Z \to$
hadrons.}, as shown in Table \ref{partialw}.

\begin{table}[h]
\begin{center}
\begin{tabular}{|l|l|l|l||l|l|}
\hline
& $\Gamma^{\rm exp}$ & $\Gamma$ & $\Gamma^{\rm SM}$ & $\Delta \Gamma$ &
$\Delta \Gamma^{\rm exp}$ \\ \hline
&&&&& \\
$Z\to e^+ e^-$ & $83.91 \pm 0.12$ & $83.452 $ & $83.480 $ & $0.028$ & $0.12$
\\
$Z \to u \bar{u}$ & & $286.060 $ & $286.100$ & $0.040$ & \\
$Z \to d \bar{d}$ & & $368.417 $ & $368.504$ & $0.087$ & \\
$Z \to$ hadrons & $1744.4 \pm 2.0$ & $1677.371 $ & $1677.712$ & $0.341$ &
2.0 \\
\hline
\end{tabular}
\caption{Partial widths $\Gamma \to f \bar{f}$ in MeV.  Note that in the
last two columns $\Delta \Gamma < \Delta \Gamma^{\rm exp}$ thus satisfying
criterion {\bf (A)} for the FSM to pass the test, despite the tree-level
predictions $\Gamma$ and $\Gamma^{\rm SM}$ being both outside experimental
bounds.}
\label{partialw}
\end{center}
\end{table}

Again as for the case of the mass,
\begin{equation}
\Delta \Gamma = \Gamma^{\rm SM} - \Gamma
\label{DelGam}
\end{equation}
represents the difference between the FSM and SM predictions, assuming again
that the loop corrections are of higher order.  This difference is in every
case less than the experimental
error, and following then the criterion {\bf (A)} we consider that the
FSM has passed the test also in this case. It is worth noting that the
accuracy of the approximate formula for the partial width far exceeds
the estimated value of the neglected parameter $\ell/k$, as explained at
the end of the last subsection.  Since the differences between the SM and
the FSM in the decay widths are already within the smaller experimental
error
in $Z$ decay, there is no need to fold in the larger error from the $W$, as
we did for the mass test.

And again, as will be shown in the last section, the difference between the
SM and FSM decreases with incresing $\zeta_S$, remaining thus within present
experimental errors for all $\zeta_S \geq 2$ TeV.

\section{New physics and the heavy partner $G$}

\begin{figure}[bth]
\centering
\includegraphics[scale=0.45]{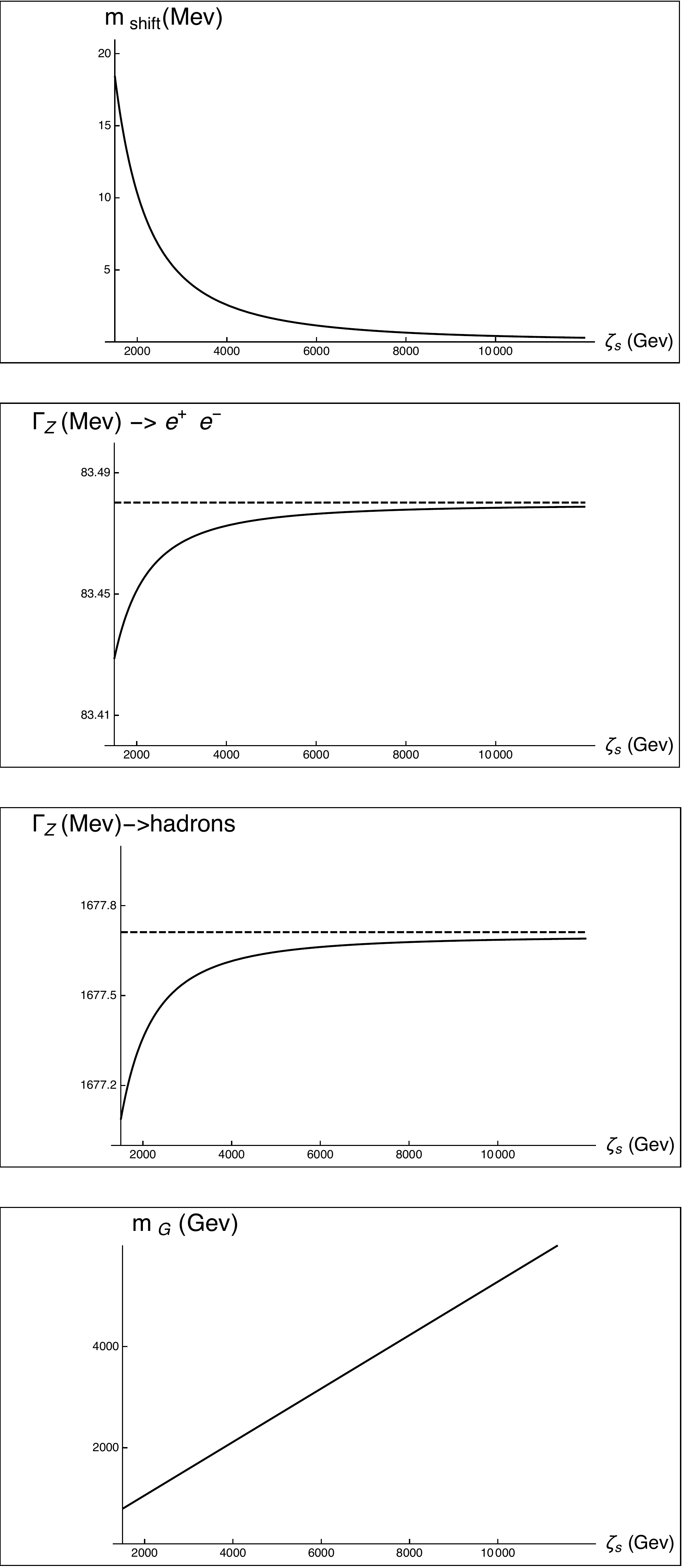}
\caption{Various plots showing $\zeta_S$ dependence---see also text}
\label{plots}
\end{figure}

The explicit formulae being known, it is straightforward to evaluate the
above deviations of FSM from SM for any value of the parameter $\zeta_S$.
In Figure \ref{plots}, we show, from top to bottom, (i) the mass shift
of the $Z$ boson $m_{\rm shift} =m_Z-m_W$, (ii) the partial width $\Gamma(Z
\to
e^+e^-)$, (iii) the partial width $\Gamma(Z \to \ {\rm hadrons})$.  We
notice
that all three deviations decrease with increasing $\zeta_S$ as anticipated,
so that having satisfied ourselves that they stay within present
experimental
bounds at $\zeta_S = 2$ TeV, we can conclude that they will do the
same for all higher values of $\zeta_S$, as claimed.

We can take a positive view and regard these deviations as new physics to be
looked for in experiment.  In connection with this, we note for possible
interest the following experimental fact.  We can deduce from equations
(\ref{zmasscompared}), (\ref{shift-delta}), (\ref{shiftexp}) that
\begin{equation}
(m_Z - m_W) < (m^{SM}_Z - m_W),
\label{mZminusmW}
\end{equation}

\begin{figure}[ht]
\centering
\includegraphics[scale=0.2]{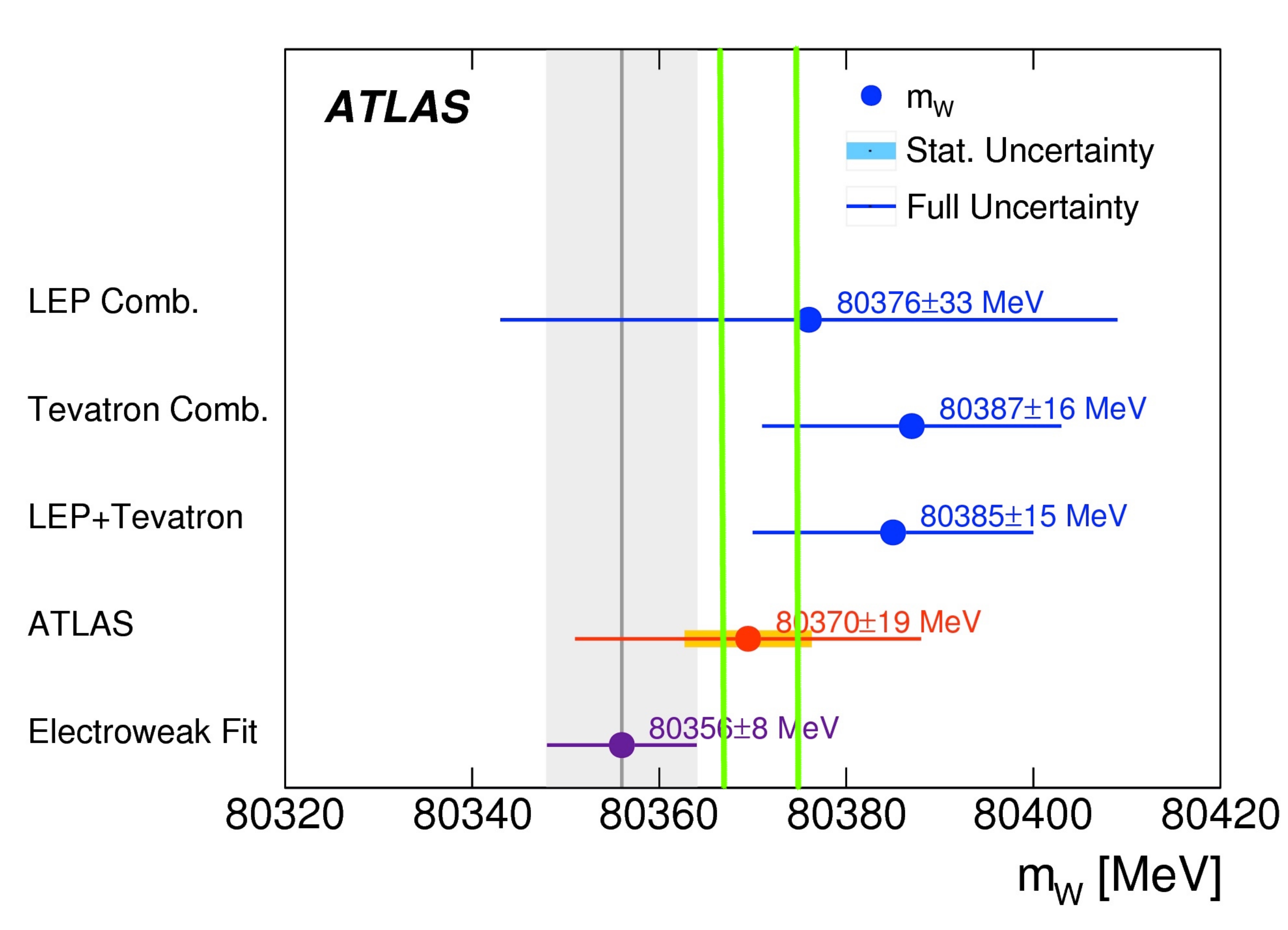}
\caption{The ATLAS measurement of the W boson mass and the combined values measured at the LEP and Tevatron colliders compared to the Standard Model prediction (mauve) and the FSM predictions (green) at $\zeta_S = 2.0$ TeV (left) and $\zeta_S = 1.5$ TeV (right).}
\label{mWfig}
\end{figure}

\newgeometry{textwidth=12cm,textheight=23cm}
\restoregeometry

a conclusion which has been checked analytically by expanding $m_Z$ in
(\ref{zmass}) to order $(\ell/k)$.  Experimentally, $m_Z$ is much better
measured (error $\sim$ 2 MeV) than $m_W$ (error $\sim$ 15 MeV), so measured
values of $m_W$ are customarily compared to the SM value as predicted by
taking $m_Z$ as input.  It is curious to note that Figure \ref{mWfig}, taken
from \cite{atlasmw}, shows that measurements at LEP, Tevatron, and LHC so
far actually all give central values for $m_W$ larger than that predicted by the
standard model, though each by only $1$-$2\ \sigma$.  Now, however, that $m_W$
in these plot should be bigger than the SM prediction is what is predicted
by (\ref{mZminusmW}) from the FSM.  So if the error can in future be reduced
and the apparent departure for the SM persists, it could be interpreted as a
point in favour of FSM.

For the decay widths also, the FSM value in Table \ref{partialw} is smaller
than the SM value because of the extra mixing, as has again been checked
analytically by expanding to order $(\ell/k)$, but here the deviations,
being
smaller, may require greater effort for their detection.

In any case, there being only the one parameter $\zeta_S$, these predicted
deviations are all correlated, so that in Figure \ref{plots} if one draws a
vertical line through the $x$-axes, one will obtain all the three different
deviations at that particular value of $\zeta_S$.  In other words, once any
one deviation is detected, the other two deviations will be predicted by FMS
in absolute terms.

Of all the predicted new physics, the most attractive to look for presumably
is the $G$ boson itself.  By virtue of mixing, it has a component in the
standard model $Z$, and can thus be produced in any reaction which produces
the $Z$ and also decay into any final states that the $Z$ does.  Hence the
$G$
boson can be searched for as a $\ell^+ \ell^-$ anomaly at the LHC in the few
TeV region, for which purpose, we would wish to know its properties.

What has been done for $Z$ in the last two sections can also be done for this
third heavy partner $G$ in the mixing complex.  Thus, the $G$ boson has a
mass given by the larger root of the quadratic equation (\ref{quadratic}),
and for our benchmark value $\zeta_s=2$ TeV, it is
\begin{equation}
m_G \approx 1057\ {\rm GeV}.
\label{Gmass}
\end{equation}
Since the running of the couplings $g_1, g_2$, and $g_3$ is small in this
range, it does not, for the present purpose, make any appreciable numerical
difference at which scale we do the calculation.

\noindent

The dependence of the $G$ mass on $\zeta_S$ is rather simple.  If one takes
the form of the discriminant given in (\ref{discrimpos}) and neglects the
last
 term proportional to $g_1^4 \sim 0.02$, one sees that the $G$ mass is
linear
in $\zeta_S$ to a very good approximation
\begin{equation}
m_G \approx \frac{\sqrt{2}}{3} \zeta_S \sqrt{(g_1^2
  +\tfrac{3}{4}g_3^2)},
\end{equation}
as is borne out in the Figure \ref{plots}.

For the decay of $G$ into a fermion-antifermion pair we can extract a
neutral current from the electroweak Lagrangian,
just as for the $Z$, and the formulae  (\ref{caandcv}),
(\ref{gamma}) and (\ref{fermi}) are analogous, except that we replace
$X_-,Y_-$ by $X_+,Y_+$,
according to the mixing matrix (\ref{mixing}); and also we have $m_G$
instead of $m_Z$.  In other words, we get the formulae:
\begin{equation}
\Gamma (G \to f \bar{f}) = \frac{m_G}{48 \pi} (c_V^2+c_A^2).
\label{Gtoff}
\end{equation}
where
\begin{eqnarray}
c_V &=& g_1X_+(-2Q+  I_3) + g_2Y_+I_3 \nonumber \\
c_A &=&g_1 X_+ I_3 + g_2 Y_+ I_3
\label{caandcvG}
\end{eqnarray}
Again for our benchmark value
$\zeta_s=2$ TeV, we get:
\begin{equation}
\Gamma (G \to e^+e^-) \sim 290\ {\rm MeV}.
\label{Gdecay}
\end{equation}

To facilitate the search for the $G$ at the LHC, we would wish to have
estimates also for its production cross section, and its total width. Both
look possible, given that the couplings of $G$ to $q \bar{q}$ and even to
the ``hidden sector" are governed by the colour gauge coupling $g_3$ already
known from perturbative QCD.  Whether it works out or not, however, is still
under investigation.

The search for $G$ will be of interest not only for its own sake. According
to \cite{cfsm}, $G$ is in the FSM one of only two known portals into the
``hidden sector''.  Being itself mostly in the hidden sector, it will, once
produced, decay preferentially into particles which are hitherto unkown to
us,
including in particular some eligible dark matter candidates.  And, any
information one can gather in that direction at present will be of paramount
significance.

\end{document}